**Protein abundances and interactions coevolve to promote functional complexes while suppressing non-specific binding.**

Muyoung Heo[1], Sergei Maslov[2], Eugene I. Shakhnovich[1]

(1) Department of Chemistry and Chemical Biology, Harvard University,
12 Oxford St., Cambridge, MA 02138, USA

(2) Department of Condensed Matter Physics and Materials Science, Brookhaven National Laboratory, Upton, NY 11973, USA

Corresponding Author: EIS, Eugene@belok.harvard.edu

**Abstract**

How do living cells achieve sufficient abundances of functional protein complexes while minimizing promiscuous non-functional interactions? Here we study this problem using a first-principle model of the cell whose phenotypic traits are directly determined from its genome through biophysical properties of protein structures and binding interactions in crowded cellular environment. The model cell includes three independent prototypical pathways, whose topologies of Protein-Protein Interaction (PPI) sub-networks are different, but whose contributions to the cell fitness are equal. Model cells evolve through genotypic mutations and phenotypic protein copy number variations. We found a strong relationship between evolved physical-chemical properties of protein interactions and their abundances due to a "frustration" effect: strengthening of functional interactions brings about hydrophobic interfaces, which make proteins prone to promiscuous binding. The balancing act is achieved by lowering concentrations of hub proteins while raising solubilities and abundances of functional monomers. Based on these principles we generated and analyzed a possible realization of the proteome-wide PPI network in yeast. In this simulation we found that high-throughput affinity capture - mass spectroscopy experiments can detect functional interactions with high fidelity only for high abundance proteins while missing most interactions for low abundance proteins.

**Introduction**

Understanding general design principles that govern biophysics and evolution of protein-protein interactions (PPI) in living cells remains elusive despite considerable effort. While strength of interactions between functional partners is undoubtedly a crucial component of a successful PPI (positive design), this factor represents only one aspect of the problem. As with many other design problems, an equally important aspect is negative design, i.e. assuring that proteins do not make undesirable interactions in crowded cellular environments. The negative design problem for PPI got some attention only recently (1, 2). Furthermore, interaction between two proteins depends not only on their binding affinity but also on their (and possibly other proteins) concentrations in living cells (2). Therefore one might expect that control of protein abundances is a third important factor in design and evolution of natural PPI. Mechanistic insights of how PPI co-evolve with protein abundances could best be gleaned from a detailed bottom up model, where biophysically realistic thermodynamic properties of proteins and their interactions in crowded cellular environments are coupled with population dynamics of their carrier organisms.

Recently we proposed a new multiscale physics-based microscopic evolutionary model of living cells (3, 4). In the model, the genome of an organism consists of several essential genes that encode simple coarse-grained model proteins. The physical-chemical properties of the model proteins, such as their thermodynamic stability and interaction with other proteins are derived directly from their genome sequences and intracellular concentrations using knowledge-based interaction potentials and statistical-mechanical rules governing protein folding and protein-protein interactions. A simple functional PPI network is postulated, and organismal fitness (or cell division rate) is presented as a simple intuitive function of concentration of functional complexes (4). While clearly quite simplified, this model provided insights into mechanisms of clonal dominance in bacterial populations and their adaptation from first principles physics-based analysis (4, 5). Here, we extend this microscopic multiscale model to study how functional PPI are achieved in co-evolution with protein abundance in living cells. We postulate a straightforward fitness function that depends on simple yet diverse functional PPI network and find that intra-cellular abundances of proteins evolve to anti-correlate with their node degrees in this network. A proteome-wide simulation, which incorporates correlations between PPI network topology, protein abundances, and interaction strengths predicted by our simple model, reproduces well the observations from high throughput Affinity Capture – Mass Spectrometry (AC-MS) experiments in yeast thus providing guidance to their interpretation.

**Results**

We designed a model cell for computer simulations, which consists of two different functional gene groups: cell division controlling genes (CDCG) and a mutation rate controlling gene (MRCG) mimicking the *mut*S protein in *Escherichia coli* and similar systems in higher organisms (see Methods). Products of CDCGs determine growth rate (fitness) as described below (Eq.(3)), while the product of MRCG determines mutation rate as in earlier study (5). All proteins can interact in the cytoplasm of model cell. Though real metabolic networks responsible for cell growth and division are very complex, we postulate a highly simplified yet diverse PPI network of CDCG as shown in Fig. 1A. Out of six CDCGs, protein product of the "first" gene is functional in a monomeric form, protein products of the "second" and "third" genes must form a heterodimer ("stable pair") to function, and protein products of the "fourth", "fifth", and "sixth" genes form a triangle PPI sub-network as shown in Fig. 1A, meaning that each protein can functionally interact by forming a heterodimer with any other protein from this sub-network (a "date triangle"). Such motifs formed by pairwise interactions of low-degree proteins with each other are common in real-life PPI networks (see ref. (6)). In this study we prohibit the formation of multi-protein complexes containing three and more simultaneously interacting proteins. Further, we posit:

1) Proteins can function only in their native conformation(s). For each protein we designate one (arbitrarily chosen) conformation as "native".

2) Protein complexes are functional only in a specific docked configuration. For each pair of proteins, which form a functional complex we designate one of their docked configuration (out of total 144 possible docked configurations of our model proteins, as explained in (4) and Methods) as functional. "Stable pair" proteins (proteins "2" and "3", k=1) have one functional surface each and participants in "date triangles" (proteins "4", "5","6", k=2) have two distinct functional surfaces each (7)) .

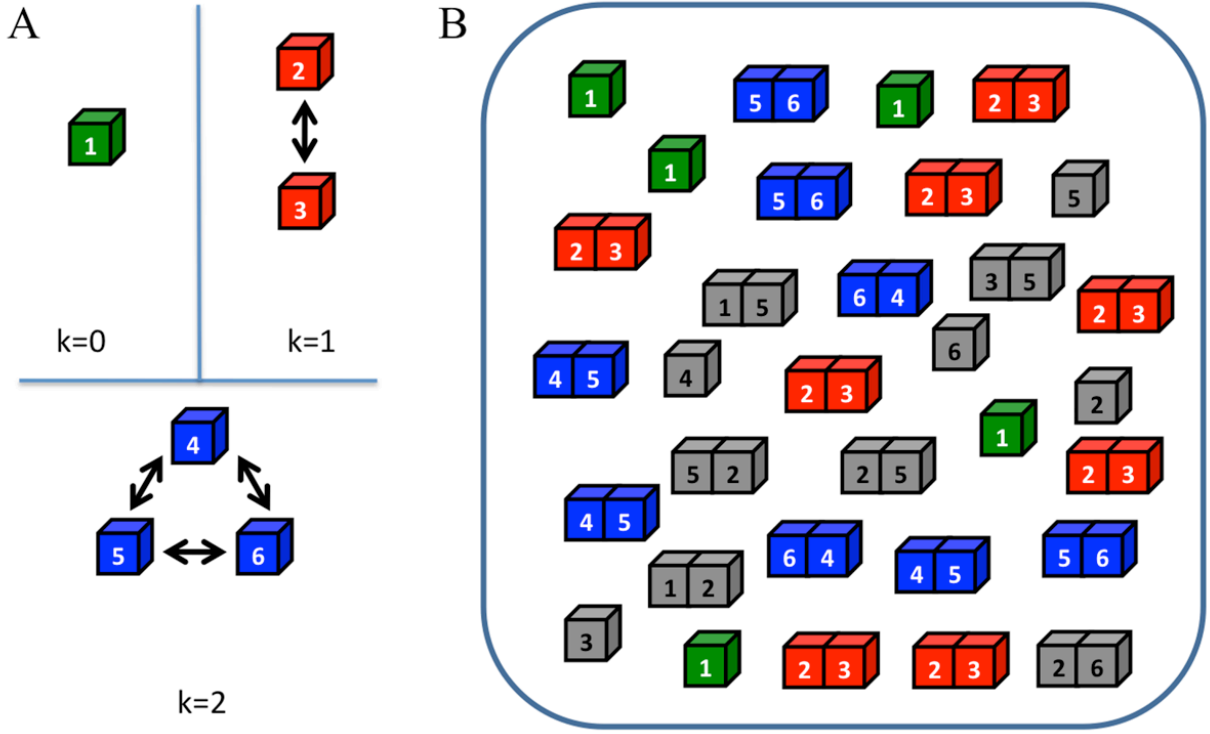

**Figure 1. A schematic diagram of the model cell.** **(A)** A model cell consists of six cell division controlling genes (CDCG) which are expressed into multiple copies of proteins. The CDCGs constitute three independent pathways with different PPI network topologies. The first protein functions in a free state (monomer, green cubes). The second and third proteins exclusively form a functional heterodimer ("stable pair") (red), but the fourth, fifth and sixth proteins circularly establish three functional heterodimers. ("date triangle", blue). **(B)** Within a cell, proteins can stay as monomers or form dimers, whose concentrations are determined by interaction energies among them through the Law of Mass Action Eqs. (S4, S5). The cubes colored as in (A) represent CDC proteins in their functional states that contribute to organism's fitness (growth rate) according to Eq. (3). Gray cubes represent proteins in their non-functional states.

Under these assumptions we define effective, i.e. *functional* concentrations of functional monomeric protein and all functional dimeric complexes:

$$G_1 = F_1 P_{nat}^1 \qquad (1)$$

where $F_1$ is total concentration of protein "1" in its monomeric form (determined from Law of Mass Action (LMA) Equations, see Ref (4) and Supplementary Text) and $P_{nat}^1$ is Boltzmann probability for this protein to be in its native state (see Methods). Functional form of "stable pair" proteins 2 and 3 and "date triangle" proteins 4,5,6 are heterodimers (the "date triangle" proteins can form more than one functional heterodimer). Effective concentrations of *functional* heterodimers of various types (i.e. 2-3, 4-5,4-6,5-6) in our model are

$$G_{ij} = D_{ij} P_{int}^{ij} P_{nat}^{i} P_{nat}^{j} \qquad (2)$$

where $D_{ij}$ is concentration of the dimeric complex between proteins $i$ and $j$ in any of the 144 docked configurations $P_{int}^{ij}$ is Boltzmann probability that proteins are docked in their functional configuration (see Ref (4) and Methods). According to the LMA $D_{ij} = \frac{F_i F_j}{K_{ij}}$ where $K_{ij}$ is the dissociation constant between proteins. The cell division rate, *i.e.* fitness of a cell is postulated to be multiplicatively proportional to all effective functional concentrations:

$$b = b_0 \frac{G_1 \cdot G_{23} \cdot \sqrt[3]{G_{45} G_{56} G_{64}}}{1 + \alpha \left( \sum_{i=1}^{7} C_i - C_0 \right)^2}, \qquad (3)$$

where $b_0$ is a base replication rate, $C_i$ is the *total* (i.e. including monomeric and dimeric forms) concentration of protein $i$, $C_0$ is a total optimal concentration for all proteins in a cell, and $\alpha$ is a control coefficient which sets the range of allowed deviations from total optimal production for all proteins. The denominator in Eq.(3) reflects the view that there is an optimal gross production level of proteins in the cell and deviations from it in either direction are penalized. Its main role is to prevent the scenario when fitness is increased due to a mere overproduction of proteins. The form of Eq.(3) is a "bottleneck"-like "AND-type" fitness function, which assumes that all CDCGs are essential for cell division. The rationale for cubic root in Eq.(3) is given in Supplementary Text.

Our first aim was to study how organisms co-evolve protein sequences and their abundances to establish functional PPI. Fig. 2A shows evolution of protein abundances. The abundance of the functionally monomeric protein (the green solid line in Fig. 2A) increases. Monomeric protein can evolve hydrophilic surfaces because the monomer does not need to have a hydrophobic binding surface shared with its functional interacting partners. (Supplementary Table I). However, abundances of functional "stable pairs" (red line) and functional "date triangles" (blue line) show quite a different trend compared with the concentration of the monomer. The total abundance of "stable pairs" proteins ($k$=1) remained approximately constant and, moreover, the total abundance of "date triangles" with $k$=2 diminished with time. In contrast to monomers, "stable pair" dimers and "date triangles" should strengthen their functional interactions by evolving strongly interacting surfaces (one surface for each "stable pair" protein and 2 surfaces for each member of "date triangle"). (see Supplementary Table I). We find that this factor limits the abundance of "stable pairs" and "date triangles" due to their enhanced propensity to form nonfunctional complexes with arbitrary partners.

In order to address the microscopic molecular mechanisms that determine optimal protein abundances, we evaluated, for each protein, the fraction of its nonspecific interactions, $ns_i$. This quantity is defined as:

$$ns_i = 1 - \frac{1}{C_i P_{nat}^i}\left(G_i + \sum_j G_{ij}\right), \qquad (4)$$

where summation is taken over all functional interactions of the protein $i$ (*i.e.* no terms in summation for protein 1, one functional partner for each of the "stable pair" proteins 2, 3 and 2 partners for "date triangle" proteins 4,5,6. The negative term in the Eq. (4) essentially is an estimate of the fraction of time that the protein spends in its monomeric state and/or participating in each of its functional interactions; naturally the rest of the time is spent participating in promiscuous non-functional interactions (PNF-PPI). The latter is defined as any interaction between proteins, which does not produce a functional complex. PNF-PPI include not only interactions between non-functional partners but also interactions between functional partners in non-functional docked states. The evolution of $ns_i$ is shown in Fig. 2B, while the evolution of functional protein interaction strengths, $P_{int}$ is shown in Fig. 2C. Initially, all proteins were designed to be stable but not necessarily soluble: they participated in many PNF-PPI (see Fig. S1). The fraction of PNF-PPI of the functional monomer ($k=0$) diminished to the lowest level as proteins evolved, apparently making its surface more hydrophilic (Supplementary Table I). On the other hand, the fractions of PNF-PPI of "stable pair" and "date triangle" proteins ($k=1$ and 2 correspondingly) still remain at higher levels. "Stable pair" proteins ($k=1$) evolved strong functional interaction, while keeping their non-functional surfaces less hydrophilic (Supplementary Table I). However "date triangle" proteins with two interaction partners evolved weaker functional PPI (Fig.2C), while becoming overall more hydrophobic than both functional monomer and "stable pair" dimer (see Supplementary Table I).

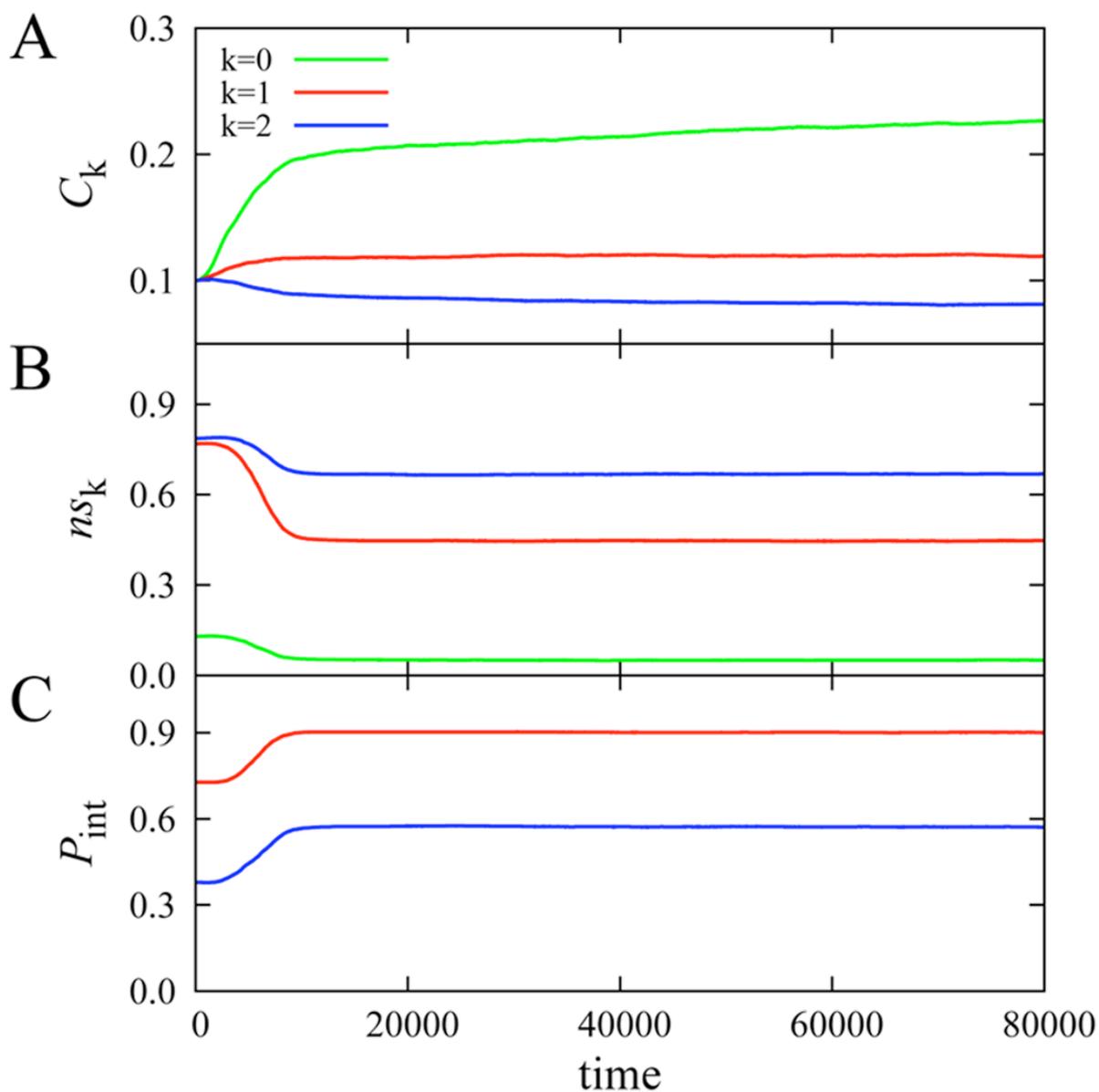

**Figure 2. Evolution of protein abundances and PPIs** after several rounds of pre-equilibration (see Fig.S1 for details). Green curves correspond to functional monomer, red curve is average over two proteins forming a "stable pair" hetero-dimer ($k=1$), and blue curve corresponds to average over three "date triangle", proteins ($k=2$). **A:** mean concentration of each protein, $C_i$. **B:** The fraction of protein material that is sequestered in non-functional interactions, $ns_i$. **C:** The strength of PPI in the functional complex, $P_{int}$, except the first protein that does not form any functional complex. All curves are ensemble averaged over 200 independent simulation runs.

To get a deeper insight into the physical origin of co-evolution between protein abundances and PPI, we investigated how relative populations of various interaction states of proteins depend on their total abundances $C_i$ (dosage sensitivity effects, Supplementary Figure 2). Functional dimers and party trimers are most susceptible to changes in their overall abundances – in fact their overproduction can cause drastic decrease in their functional concentrations. We also note that loss of functional concentrations of dimers and party trimers occurred to a considerable extent due to formation of homodimers, in line with the analysis in (8).

Functional surfaces of proteins evolved in our model are enriched in several hydrophobic amino acids. This model finding agrees well with the analyses of PPI interfaces of real proteins (9-11), which also suggest that hydrophobic interactions are the dominant force behind functional PPI (11, 12). Figure 3 compares amino acid composition on functional PPI interfaces of model and real proteins. Quite remarkably, our simple model correctly captures all six amino acid types, which are enriched *in conservative clusters* on PPI interfaces (13) (except swap between Aspartic and Glutamic acids, which such simple potential apparently cannot distinguish between). Highly significant correlation between model and real propensities for all 20 amino acids (correlation coefficient is 0.6129 and p-value is 0.0041) suggests that our model and its knowledge-based potential, despite their simplicity, capture essential aspects of physical chemistry of PPI.

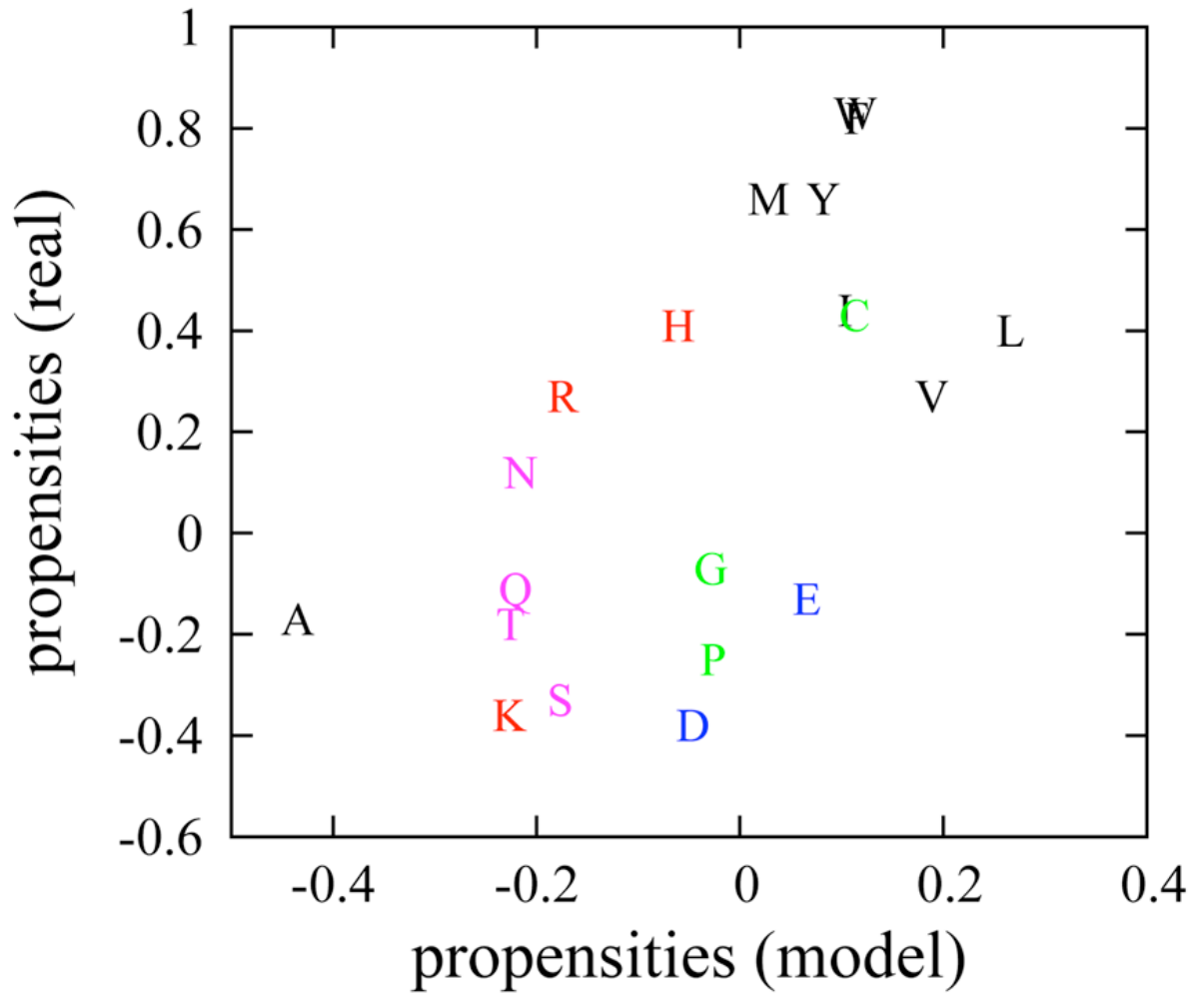

**Figure 3. Scatter plot between amino acid propensities on functional interfaces of model and real proteins.** We calculated the propensities for all model proteins from protein orthologs from 152 representative strains as described in Eq. (S6). The propensities for real proteins are obtained from Table 2 of ref. (10). The color scheme is as follows: hydrophobic (black), positively charged (red), negatively charged (blue), uncharged polar (cyan), and remaining amino acids (green).

In summary, our simple model predicts that: 1) Abundance of a protein in cytoplasm is negatively correlated with the number of its functional interaction partners (Fig.4A); 2) Strength of functional interactions of a protein is also negatively correlated with its node degree in the PPI network (Fig.2C); 3) Less abundant proteins engage in stronger PNF-PPI (see Fig.4B).

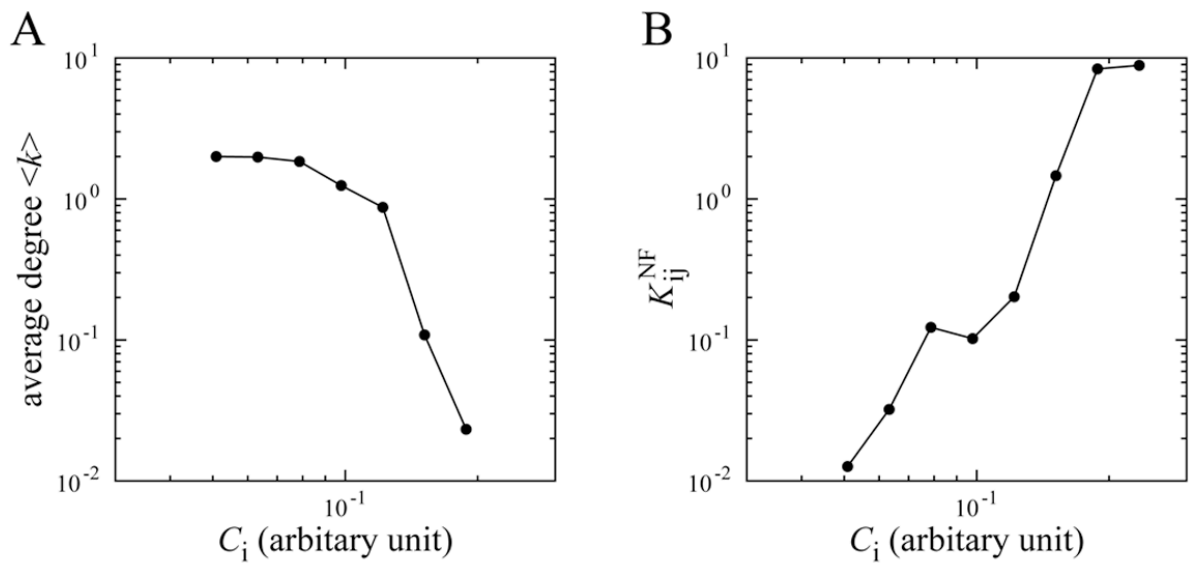

**Figure 4. The node degree in the functional PPI network and the strength of PNF-PPI negatively correlate with protein abundance.** Both the average degree $<k>$ in the functional PPI network (A) and the dissociation constants of PNF-PPI complexes, $K_{ij}^{NF}$ which is inversely proportional to the strength of PNF-PPI (B) are plotted as function of protein abundance, $C_i$.

Now we wish to test these predictions. This is not an easy task because interactomes reported in high-throughput experiments may be different from real ones due to significant fraction of false positives and missed weak functional interactions: PPI networks reported by various techniques differ greatly between techniques and experimental realizations (14). Furthermore, whole-proteome measurements of binding affinities for functional and PNF-PPI are not available. Therefore we developed the following strategy. First, we designed a reference, "true" Baker Yeast interactome, which exhibits correlations observed in the simple model. Next, we "experimentally" study this interactome using a computational counterpart of the Affinity-Capture Mass-Spec (AC-MS) PPI experiments to determine the "apparent" interactome, which might differ from the "true" one. Finally we compare the "apparent" interactome obtained computationally from the underlying "true" one with the interactome obtained in real AC-MS experiments to determine whether experimental data bear signatures of the correlations predicted from simple exact model.

We built a "true" Baker's Yeast interactome for its 3,868 proteins, whose intracellular abundances are known from experiment (15) by rewiring the published PPI network obtained in AC-MS experiments (16) to preserve its scale free character (see Figure S3) and to introduce anti-correlations between node degree and abundance as predicted by the model (see Fig.5A).

Dissociation constants of functional binary protein complexes $K_{ij}^{F}$ were assigned to reflect the negative correlations between node degree and affinity of functional complexes as found in the simple model:

$$K_{ij}^{F} = 0.01\exp\{1.5(k_i + k_j)\} \qquad (5)$$

Dissociation constants for PNF-PPI between all proteins were assigned to positively correlate with evolved abundances as predicted by the model (see Fig 4B):

$$K_{ij}^{NF} = 15\cdot\max(C_i, C_j), \qquad (6)$$

By solving 3,868 coupled nonlinear LMA equations we obtained all possible binary complex concentrations, $D_{ij}$ for the designed reference interactome. Then we mimic the AC-MS experiments by ''capturing'' only complexes whose concentration exceeds a certain "detection threshold", *i.e.* $D_{ij}/C_i \geq THR$. Here $C_i$ is the concentration of the "bait" protein and the threshold emulates finite sampling of captured complexes by mass spectroscopy. By varying the detection threshold we can approximately mimic the stringency of the detection of interactions in the AC-MS experiments by the criterion $MS \geq w$ where $w$ is the number of times an interaction is reproduced in independent AC-MS experiments.

The model counterpart of the $MS \geq 1$ interactions (low $THR=1/400$) shows an almost monotonic positive dependence of the averaged detected node degree, $\langle k \rangle$ on protein abundance except for highly abundant proteins (Fig. 5A, black line), while the model counterpart of the more stringent $MS \geq 3$ dataset (higher detection threshold $THR=1/20$) shows a non-monotonic behavior with highest $\langle k \rangle$ corresponding to proteins of medium abundance (Fig.5A, red line). Strikingly, independent of the threshold the "apparent" node degrees of low abundance proteins are much lower than their degrees in the "true" functional PPI network as most functional interactions for these proteins are missed. The probability to detect functional PPI increases drastically with protein abundance (Fig.5B). On the other hand, for high values of threshold THR "true" and "apparent" PPIs of highly abundant proteins exactly match each other corresponding to the set of highly reproducible ($MS \geq 3$) interactions, (Fig.5A) while lower values of $THR$ (or $MS \geq 1$ dataset) still include many false-positive PPI even for high abundance proteins (see Fig.5C). As regards false positives (i.e. PNF-PPI) in AC-MS experiments many of them are detected for highly abundant proteins at low detection threshold (i.e. $w \geq 1$) and are eliminated for all proteins regardless of abundance at a more stringent detection threshold (corresponding to $w \geq 3$ or greater). (Fig. 5C).

We compared the predictions of our model shown in Fig. 5A with large-scale proteomics data on *S. cerevisiae* shown in Fig. 5D. We used PPIs marked as "AC–MS" in the BioGRID database (16, 17) and protein copy numbers experimentally measured (15) under normal (rich medium) conditions. Fig. 5D plots the average degree $\langle k \rangle$ vs. protein copy numbers for each of two datasets extracted from BioGRID: all MS-detected interactions (MS$\geq$1, black symbols), and interactions reproduced in three or more independent experiments (MS$\geq$3, red symbols). Similar to the yeast proteome model, the MS$\geq$1 and MS$\geq$3 data exhibit different trends in $\langle k \rangle$ for proteins of above $C > 2 \times 10^4$ copies/cell. Whereas in the MS$\geq$1 dataset $\langle k \rangle$ systematically increases with concentration until high copy number range, in the MS$\geq$3 dataset $\langle k \rangle$ reaches maximum value $\approx 2$ at protein concentrations around $2 \times 10^4$ copies/cell and then starts to systematically decrease with *C*, exactly as found for the ''true'' model proteome in which correlations predicted by the simple model are built in.

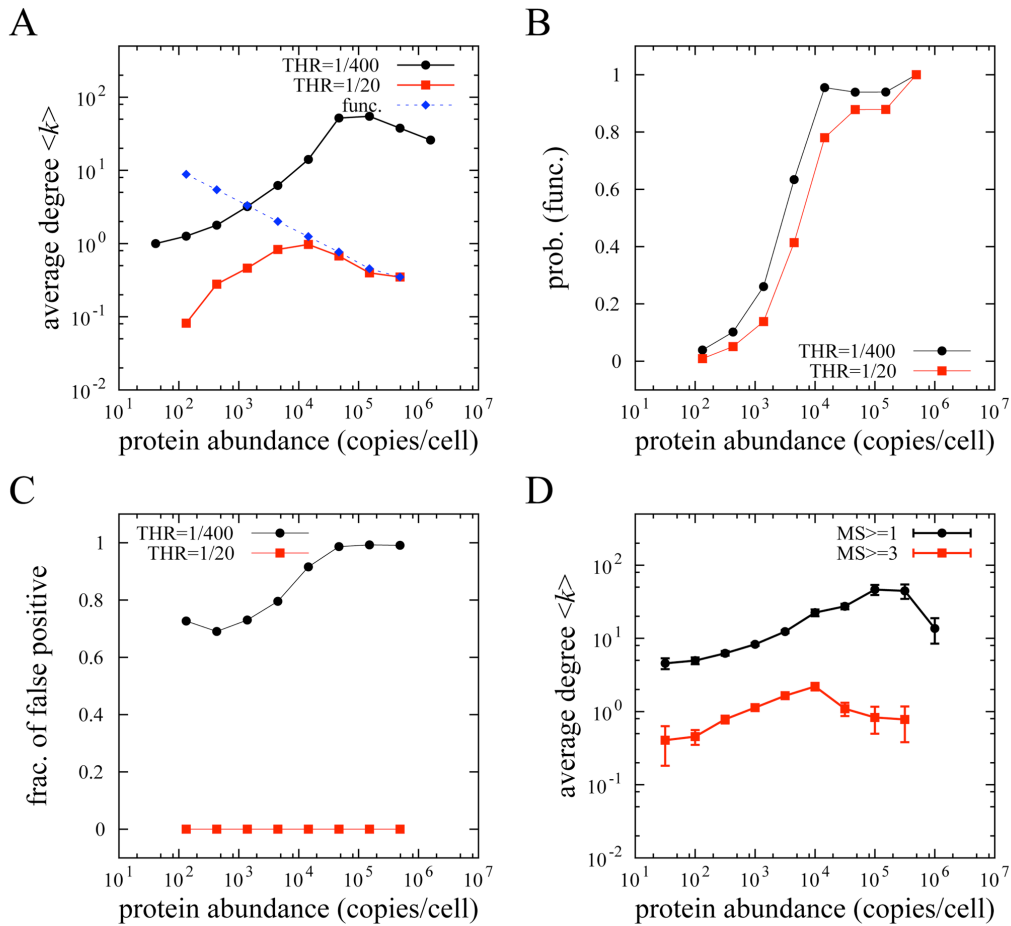

**Figure 5. System-wide proteomics simulation of PPI detection and comparison with AC-MS high throughput experiments. (A) Simulated "AC-MS" type of experiment in our**

**model.** We "designed" a set of 6228 functional interactions among 3868 proteins and assigned dissociation constants to all PPI as described in Eqs. (5,6). Blue dashed line represents the average node degree of designed "true" PPI and black and red solid lines correspond to the node degrees of "captured" PPI networks in our proteomics model at different values of detection threshold. **(B)** The fractions of functional PPIs out of all "captured" PPI in our simulation at low (black) and high (red) thresholds are plotted as a function of protein abundance. **(C) The fraction of detected PNF-PPI** out of all ''captured'' PPI. **(D) The average degree of a protein in the *S. cerevisiae* PPI network vs. protein abundance.** Black symbols correspond to all ~28,800 AC-MS labeled interactions in the BioGRID database, while red symbols correspond to ~2,600 highly reproducible interactions confirmed in three or more independent experiments.

**Discussion**

In this work we used a multiscale first-principle model of living cells to investigate the complex relationship among functional PPI, PNF-PPI, and the evolution of growth-optimal protein abundances. Despite its simplicity the model allows a microscopic *ab initio* approach to address these complex and interrelated issues. Unlike traditional population genetics models here we do not make any *a priori* assumptions of which changes are beneficial and which ones are not. Rather we base our model on a biologically intuitive genotype-phenotype relationship Eq. (3), which posits that growth rate depends on biologically functional concentrations of key enzymes (or multi-enzyme complexes), which make metabolites that are necessary for cell growth and division. In support of this view the high-throughput data of Botstein and coworkers shows that for a significant fraction of proteins their expression levels are indeed correlated with growth rates (18, 19). Overall one should expect that for enzymes whose substrate concentrations in living cells exceed their $K_M$, the turnover rates of their metabolites would be proportional to their concentrations, affecting fitness (growth rates) of carrier organisms as suggested by our genotype-phenotype relationship in Eq. (3).

Our findings provide a general framework for understanding the physical factors determining protein abundances in living cells. We found that functional monomers evolved largely hydrophilic surfaces, which allowed their production level to increase with apparent fitness benefit and minimal cost due to PNF-PPI. This finding is consistent with the observation that in *E. coli* more abundant proteins are less hydrophobic (20). In contrast intracellular copy numbers of proteins participating in multiple functional PPI evolve under a peculiar physical constraint: such proteins have to evolve hydrophobic interacting surfaces to provide strong functional PPI, as found in our simulations and also established in several statistical analyses of known functional complexes (21-23). However the same hydrophobic surfaces contribute to promiscuous non-functional interactions. This "frustration" between functional and non-functional interactions is resolved by limiting effective concentrations of "stable pairs" and "date triangles" in our model cells and weakening of their functional PPI. Recent computational analysis of PPI energetics confirmed this prediction by demonstrating that proteins which have more functional partners in the PPI network have weaker functional interactions (24). An interesting possibility to overcome this frustration effect is to keep sequences of some proteins, which have multiple interaction partners, hydrophilic by making these proteins intrinsically disordered as has been indeed observed (25).

More generally, our main finding is that protein abundance, being directly dependent on physical properties of proteins such as their participation in PPI network, may be under tight evolutionary control. The implication is that as long as PPI network is conserved between species, abundance should be conserved between orthologous proteins as well (e.g. more conserved than their mRNA levels). This is indeed the case (26).

Our high-throughput computational analysis of functional and PNF-PPI in proteome of *S.cerevisae* provided an insight into inner working of AC-MS experiments and a guidance to their interpretation. It appears that functional PPI of highly abundant proteins (copy numbers in cytoplasm exceeding $2 \times 10^4$) are recovered quite well when an interaction is reproduced in multiple independent AC-MS experiments. The situation is not so rosy for low abundance proteins since large fraction of their functional interactions is not captured in AC-MS data at any detection threshold. Lowering the detection threshold somewhat increases the fraction of detected functional interactions for medium abundance proteins but at a cost of mixing in even larger number of non-specific interactions.

Our model while capturing many realistic biophysical aspects of proteins and their interactions is still minimalistic as it focuses on the relation of the physical properties of proteins to cell's fitness and disregards certain aspects of their functional behavior in living cells. One possible limitation is that our model of PPI interfaces and interaction potentials may be too simple to capture complex aspects of PPI specificity such as steric complementarity (lock and key), conformational change and highly specific directional interactions. However a thorough analysis of PPI energetic and structural data by many groups (reviewed in (11, 12)) shows that: 1) The majority (over 90%) of PPI interfaces are planar 2) the same majority of interfaces exhibit very little if any conformational change and 3) the major contribution to stability of PPI comes from hydrophobic interactions (mostly aromatic but aliphatic as well) as seen from alanine scan experiments and interface composition analyses. However there are known cases (e.g. involving intrinsically disordered proteins (25)) when conformational changes leading to formation of PPI interfaces are apparent, and our model does not apply to these situations. To that end our predictions are of intrinsically statistical nature. Nevertheless, the physical mechanisms discussed here are common to most proteins in the cell and we expect that interplay between functional and non-functional interactions prove to be an important factor determining evolution of protein abundance.

**Methods**

**Protein structure and interactions**

Our model cells carry explicit genome, which is translated into 7 different proteins: a functional monomer, two "stable pair" proteins, three members of the "date triangle", and a homodimeric protein defining the mutation rate of the cell. For simple and exact calculations, proteins are modeled to have 27 amino acid residues and to fold into 3x3x3 lattice structures (27). Only amino acids occupying neighboring sites on the lattice can interact and the interaction energy depends on amino acid types according to the Miyazawa-Jernigan potential (28) both for intra- and inter-molecular interactions. For fast computations of thermodynamic properties we selected 10,000 out of all possible 103,346 maximally compact structures (27) as our structural ensemble. This representative ensemble was carefully selected to avoid possible biases (4). As a measure of protein stability, we use the Boltzmann probability, $P_{nat}$, that a protein folds into its native structure.

$$P_{nat} = \frac{\exp[-E_0/T]}{\sum_{i=1}^{10000} \exp[-E_i/T]} \tag{7}$$

where $E_0$ is the energy of the native structure – a conformation, which is *a priori* designated as the functional form of the protein, and $T$ is the environmental temperature in dimensionless arbitrary energy units.

We use the rigid docking model for protein-protein interactions. Because each 3x3x3 compact structure has 6 binding surfaces with 4 rotational symmetries, a pair of proteins has 144 binding modes. For each protein that participates in a given functional PPI one surface is *a priori* designated as "functionally interacting" and one heterodimeric configuration/orientation is *a priori* designated as the functional binding mode. Proteins 4,5,6 forming "date triangles" have two binding surfaces each. The Boltzmann probability, $P_{int}^{ij}$ that two proteins forming a binary complex interact in their functional binding mode (out of 144 possible ones) and the binding constant, $K_{ij}$ between proteins *i* and *j* are evaluated as follows:

$$P_{int}^{ij} = \frac{\exp[-E_f^{ij}/T]}{\sum_{k=1}^{144} \exp[-E_k^{ij}/T]}, \qquad K_{ij} = \frac{1}{\sum_{k=1}^{144} \exp[-E_k^{ij}/T]} \tag{8}$$

where $E_f^{ij}$ and $E_k^{ij}$ are respectively the interaction energy in the functional binding mode (where applicable) and the interaction energy of *k*-th binding mode out of 144 possible pairs of sides and mutual orientations between the proteins *i* and *j*.

**Simulation**

The initial sequences of proteins were designed (29, 30) to have high stabilities ($P_{nat}^i > 0.8$) and their native structures were assigned at this stage and fixed throughout the simulations. Initially, 500 identical cells were seeded in the population and started to divide at rate of $b$ given by Eq. (3). In order for both genotypic and phenotypic traits of organisms to be transferred to offspring, a cell division was designed to generate two daughter cells, whose genomes and protein production levels, $C_i$s are identical to those of their mother cell except genetic mutations that arise upon division at the rate of $m$ per gene per replication as following:

$$m = m_0 \left(1 - \frac{G_{77}}{G_{77}^0}\right), \quad (9)$$

where $G_{77}^0$ is the initial functional concentration of mismatch repair homodimers of the seventh protein. At each time step, we stochastically change the protein production level, $C_i$ with rate of $r = 0.01$ to implicitly model epigenetic variation of gene expression (5, 31).

$$C_i^{new} = C_i^{old}(1 + \varepsilon), \quad (10)$$

where $C_i^{old}$ and $C_i^{new}$ are the old and new expression levels of protein product of $i$-th gene, and $\varepsilon$ is the change parameter which follows a Gaussian distribution whose mean and standard deviation are 0 and 0.1, respectively.

The population evolved in the chemostat regime: the total population size was randomly trimmed down to the maximum population size of 5000, when it exceeded the maximum size. The optimal total concentration of all proteins, $C_0$, is set to 0.7. The death rate, $d$, of cells is fixed at 0.005 per time units, and the parameter $b_0$ is adjusted to set the initial birth rate to fixed death rate ($b=d$). The control coefficient $\alpha$ in Eq. (3) is set to 100. 200 independent simulations are carried out at each condition to obtain the ensemble averaged evolutionary dynamics pathways.

**Acknowledgements:**

Work at Brookhaven National Laboratory was carried out under Contract No. DE-AC02-98CH10886, Division of Material Science, US Department of Energy. Work at Harvard is supported by the NIH.

**Supplementary Text**

**Concentration dependence of fitness function: why cubic root?**

The stoichiometric balance of protein concentrations in our model is given by the conservation equation:

$$C_i = F_i + \sum_{j=1}^{7} F_{ij} \qquad (S1)$$

For simplicity consider a well-evolved organisms where functional interactions dominate, i.e. $K_{ij}^{F} \ll K_{ij}^{NF}$. Then most proteins are in their functional form and we get:

$$\begin{aligned}
F_1 &\approx C_1 \\
F_{23} &\approx C_2 \approx C_3 \\
F_{45} + F_{64} &\approx C_4 \qquad (S2) \\
F_{45} + F_{56} &\approx C_5 \\
F_{46} + F_{56} &\approx C_6
\end{aligned}$$

In this regime contribution to fitness function from dimers and date trimers are as follows:

$$\begin{aligned}
F_{23} &= \frac{1}{2}(C_2 + C_3) \\
F_{45} F_{56} F_{64} &= \frac{1}{8}(C_4 + C_5 - C_6)(-C_4 + C_5 + C_6)(C_4 - C_5 + C_6)
\end{aligned} \qquad (S3)$$

which explains why cubic root in fitness function Eq.(3) of Main Text is necessary to avoid bias which a'priori favors one type of complexes over the other.

**Solution for the Law of Mass Action (LMA) equations**

For simplicity, proteins are modeled to form only monomers or dimers and all the higher order protein complexes are ignored in this work. The monomer concentrations of proteins, $F_i$ were determined by solving the following seven coupled nonlinear equations of LMA (3, 4):

$$F_i = \frac{C_i}{1+\sum_{j=1}^{N}\frac{F_j}{K_{ij}}} \text{ for } i=1,2,\cdots,N, \tag{S4}$$

where $N$ is the number of proteins in the system ($N=7$ for *ab initio* model and $N=3868$ for proteomics simulation model) and $K_{ij}$ defined in Eq. (8) (for *ab initio* model) and Eq. (5, 6) (for proteomics simulation model) of the text is the average dissociation constant of all possible interactions between proteins $i$ and $j$. The concentration $D_{ij}$ of dimer complex between any pair of proteins are then given by the following LMA relations:

$$D_{ij} = \frac{F_i F_j}{K_{ij}}. \tag{S5}$$

We solved seven coupled nonlinear equations of LMA using the iteration method of (3, 4): one calculates the first iteration of $F_i$ by substituting $C_j$ for $F_j$ in the right hand side of the Eq. (S1). Each new iteration of $F_i$ is then plugged in the right hand side of the Eq. (S1). The iterations are repeated until the maximum relative deviation of the new values of $F_i$ from the old ones drops below $10^{-6}$.

**Hydrophobicities of evolved proteins**

To characterize the hydrophobicity of the amino acids in simulations we note that 20*20 matrix of Miyazawa-Jernigan potentials allow spectral decomposition with one type eigenvalue, (5) i.e. an element of the matrix describing interaction energy between amino acids i and j can be presented as: $E_{ij} = E_0 + \lambda q_i q_j$ where $q_i$ is an effective hydrophobicity index of an amino acid of type i which ranges from $q_{min} \approx 0.125$ (most hydrophilic, K) to $q_{max} \approx 0.333$ (most hydrophobic, F). We rescaled the hydrophobicity scale to fall into (0,1) interval: $\tilde{q}_i = \frac{q_i - q_{min}}{q_{max} - q_{min}}$. These values are presented in Table I.

**Propensities of 20 amino acids constituting functional interfaces**

We defined the propensity, $\Pr_a$ to find an amino acid type $a$ in functional interfaces as follows.

$$\Pr_a = \ln \frac{p_a}{p_a^0}, \qquad (S6)$$

where $p_a$ and $p_a^0$ are the probabilities to find an amino acid type $a$ in sequence regions corresponding to functional interfaces and all sequence, respectively.

**PPI and protein abundance data for *S. cerevisiae***

We downloaded the genome-wide PPI network in baker's yeast *S. cerevisiae* from the BioGRID database (6, 7) and extracted all bait-to-prey pairs of interacting proteins detected by the affinity capture followed by mass spectrometry technique (designated as "Affinity Capture-MS" in the database). A pair of interacting proteins was then included in our "MS$\geq w$" dataset if it was confirmed by at least *w* independent mass spectrometric experiments. We also obtained the protein expression levels of yeast proteins measured by Ghaemmaghami *el. al* (8). All proteins are classified with respect to their protein copy numbers using log bins. Fig. 5D shows plots the average degree of all proteins in the same concentration bin in different MS$\geq w$ datasets: *w*=1 (black symbols) and 3 (red symbols).

**Supplementary Figures**

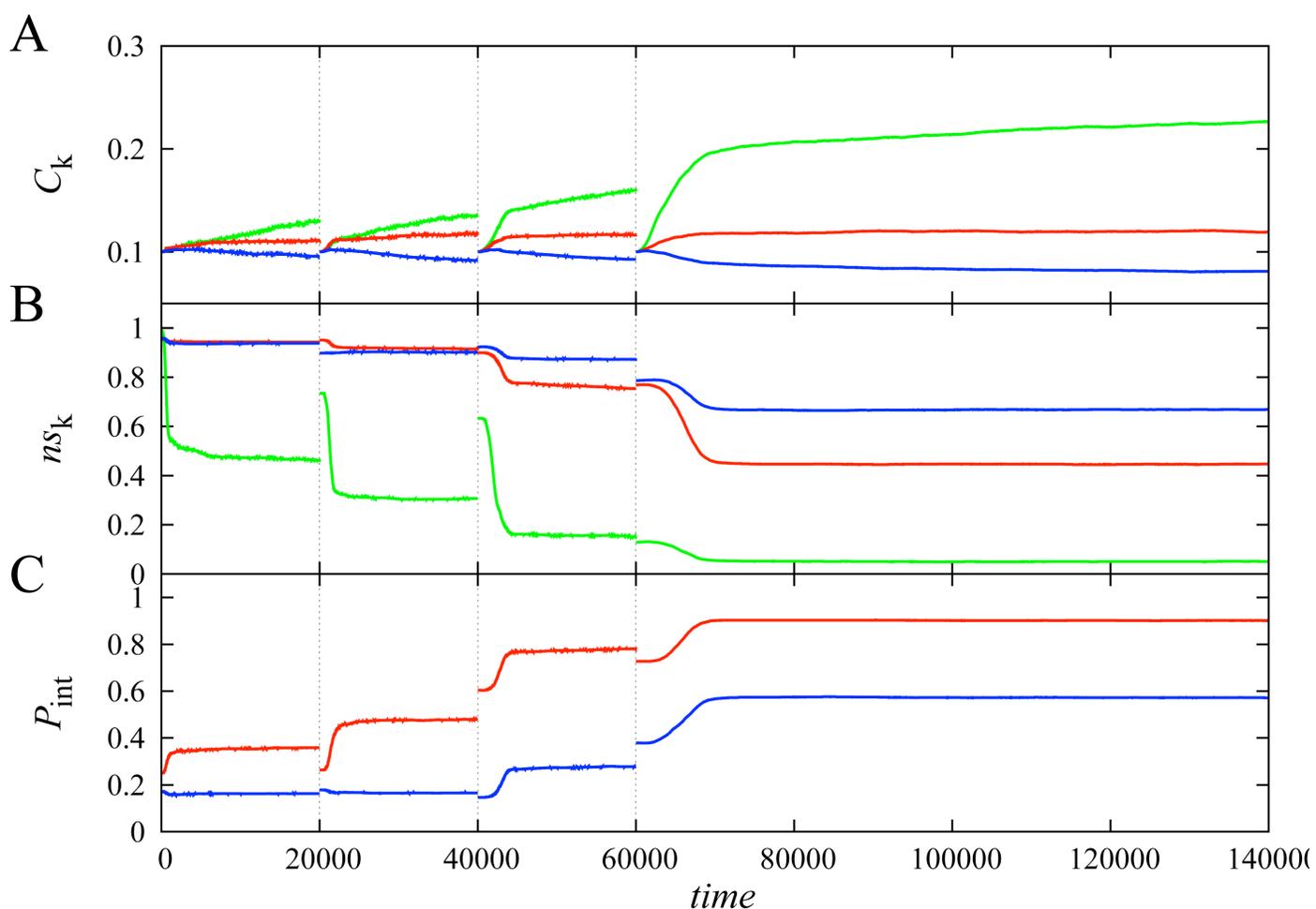

**Figure S1. Evolution of protein abundances, functional and nonfunctional protein-protein interactions.** The curves in each panel represent total protein concentrations (top), fractional concentrations of a protein forming nonfunctional complexes (middle), and the probability to form a functional PPI complex (bottom). The color codes represent functional monomer (protein 1, green), "stable pair" having one functional partner (protein 2 and 3, red), and "date triangle" with two functional partners (protein 4, 5, and 6, blue). We designed initial sequences of 6 cell division controlling genes (CDCG) to have highly stable structures ($P_{nat} > 0.8$) without regard for solubility of their surfaces, which resulted in mostly promiscuous nonfunctional binding of initial proteins with one another. Our population dynamics simulation consists of two parts: the first three consecutive simulations to equilibrate proteins to have proper functional interfaces depending on their functional requirements (20000 simulation time step each up to $t$=60000) and the last long time production run simulation from $t$=60000 to $t$=140000, which corresponds to the simulation data presented in Fig. 2 in main text. The vertical dotted lines partition different rounds of simulations. The seeding genome for the next round of simulation is randomly picked up out of the evolved organisms in the previous round of simulation, (roughly mimicking serial passage experiments) which explains the discontinuities at $t$=20000, 40000, and 60000. In all cases, the fraction of nonfunctional interactions of the functional monomer most drastically drops at the early

stages of each round of simulation. On the other hand, the variations of nonfunctional and functional interactions of "date triangle" proteins are smaller than those of "stable pair" proteins. We averaged the curves over 100 different simulations for the first three rounds of simulations and 200 different simulations for the last round of simulation.

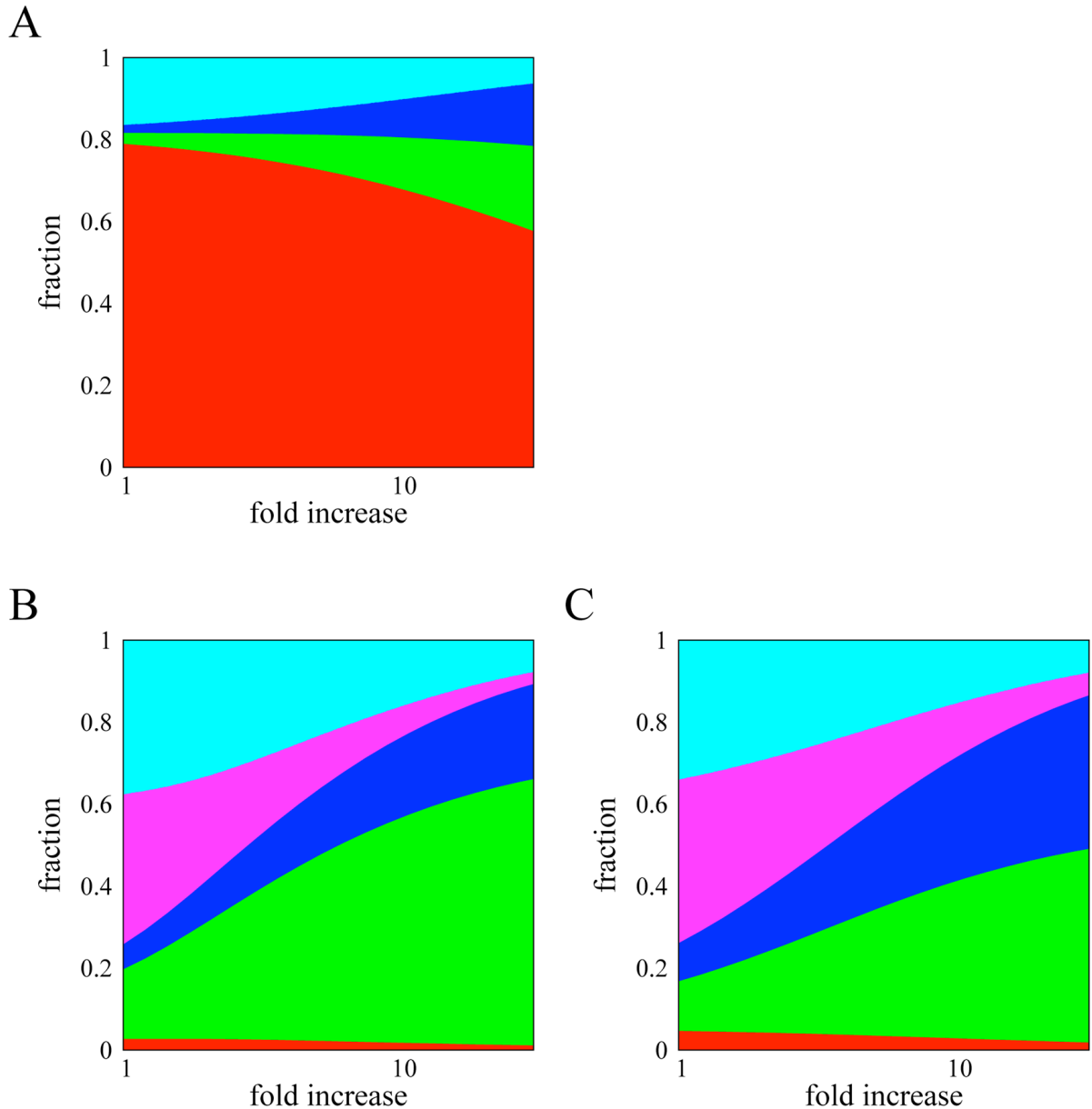

**Figure S2. Effect of dosage increase on the formation of various complexes.** Colors denote various types of states of a protein: monomer (red), homodimer in head-to-head form which shares the same binding interface (green), homodimer in head-to-tail form where two participants use different binding interfaces (blue), functional heterodimer (magenta), and promiscuous complexes with a random partner (cyan). The width of each strip corresponds to the fraction of proteins in corresponding

states/complexes in the cytoplasm of the model cell. The X-axis quantifies the level of overexpression relative to the wildtype (evolved) concentration (A) functional monomer protein. (B) "stable pair" functional dimer proteins (C) functional dimer proteins involved in the "date triangle".

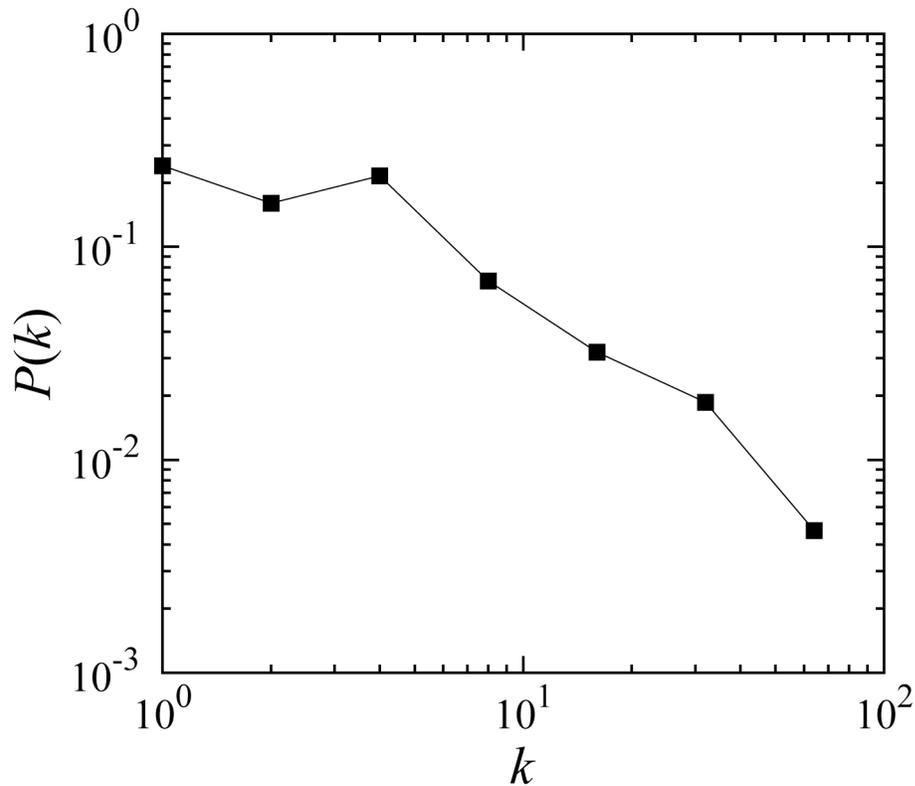

**Figure S3.** The probability, *P(k)* to find a protein having node degree *k*. The artificially made "true" PPI network for 3868 proteins of Baker's yeast retains the scale-free property of the original one.

**Supplementary References**

**Supplementary Table**

| The number of PPI partners | Hydrophobicity per residue | | |
|:---:|:---:|:---:|:---:|
| | Functional interface | Non-binding region | Overall sequence |
| k=0 | N/A | 0.29±0.02 | 0.29±0.02 |
| k=1 | 0.50±0.02 | 0.29±0.03 | 0.36±0.02 |
| k=2 | 0.49±0.03 | 0.30±0.05 | 0.43±0.02 |

Supplementary Table I. **Hydrophobicity of evolved proteins.** Average and standard deviations of relative normalized hydrophobicity per residue of each sequence region. The relative normalized hydrophobicity scales from 0 (most hydrophilic) to 1 (most hydrophobic). Averages and standard deviations are calculated over protein orthologs from 152 representative strains as described in Supplementary Text.